**Developing Mathematics for Insight into Sensorimotor Neurobiology**

Gin McCollum
Neuro-otology Department
Legacy Research Center
1225 NE 2nd Avenue
Portland, Oregon 97232
USA
mccollum@ohsu.edu

021108

**Abstract**

This paper summarizes a research program to express the organization of sensorimotor control by specifying physiological states and the conditions for transitions among them.  By a slight change in standard notation, conditional dynamics provides a moving spotlight, focussing on salient subspaces within a high-dimensional space.  This mathematical approach serves as a window on the organization of sensorimotor neurobiology.  The intertwined efforts to express the intrinsic organization of neurobiology and to clarify it mathematically are yielding a mathematical structure that is growing on fertile empirical ground.

A great body of mathematics has been developed to express the organization of the physical world. In turn, the mathematics cantilevers ideas of the physical world into new conceptual spaces, suggesting questions about the physical world. For example, the mathematics of quantum structures has grown on the empirical foundation of quantum mechanics and provides a continuing dialogue about foundational concepts in physics. In the neurobiology of sensorimotor control, there is now sufficient empirical foundation to develop mathematics characterizing the intricate organization of sensorimotor behavior. This paper briefly outlines a project developing such mathematics.

*Discrete/Continuous Structure in Sensorimotor Neurobiology*

A combination of discrete and continuous properties occurs in the neurobiology of sensorimotor control. For example, sleeping and waking are discrete states analogous to the discrete states of an electron in an atom (McCollum, 1999b). Continuous dynamics give complementary descriptions of these states, both neural and physical. In the mathematics we are developing, the discrete structure is currently of most interest.

This discrete structure characterizes the relationships and transitions between states (McCollum, 1994b, 2002). The responses of an organism depend on its state, in the sense that the state narrows the perceptual and motor choices. For example, speaking a certain language is a state of the human nervous system. A person who is speaking English will hear as "nine" the same sounds that a person in the state of speaking German will hear as "nein". Transitions among states are central to the mathematics we are developing.

The discrete/continuous structure of the mathematics we are developing in theoretical neurobiology provides a natural relationship to the mathematics of quantum structures. The mathematics of both quantum mechanics and neurobiology bridges various areas of mathematics, especially algebra and analysis. I appreciate the inspiration and critique offered by the quantum structures community over the years. Discrete measurements are central in the mathematics of quantum mechanics. Just as the outcomes of measurements in quantum mechanics depend on the ordering, so an organism's responses depend on ordering. In quantum mechanics, however, the physicist's choice of measurement is outside the theory. In the approach we have taken in sensorimotor neurobiology, the state of the system and the conditions for transitioning between states are central; the state of the system introduces sensorimotor constraints analogous to choice of measurement (McCollum, 1994a).

*Sensorimotor Space*

The sensorimotor states we typically study integrate many degrees of freedom. For example, trunk movements, head movements, and eye movements form a coordinated dynamical system of many degrees of freedom. The neural centers mediating the movements add many more: several for each neuron involved.

It is often possible to reduce the number of degrees of freedom by choosing the most salient ones. For example, major features of postural adjustments (Nashner & McCollum, 1985;

McCollum et al, 1985) and the sit-to-stand movement (Roberts & McCollum, 1996) can be characterized using just two degrees of freedom. A natural state of the system can be specified as a *region*, or subset of a subspace of a high-dimensional space. For example, the state of walking involves each leg following a cycle in a plane within the three-dimensional space of ankle, knee, and hip joint angles (Borghese et al 1996; Bianchi et al 1998). Because sensorimotor control is within a region rather than exact, the cross-section of the constraining cycle is two-dimensional. A wide variety of neural and behavioral states can be specified in this way.

Over an extended movement sequence, the salient sets of degrees of freedom typically change. For example, in walking, ankle variables are salient during heel strike, hip variables are salient in matching the two legs' strides, and ankle, knee, and hip coordination over a single leg is salient for obstacle avoidance. Each segment and phase of walking affects the others, even though they may occur mostly in different subspaces of the overall movement space. When sensory control is added, there are even more variables and transitions among salient variables. Thus, the sensorimotor space is typically of high dimension for a sensorimotor sequence.

*Variable Subspaces within a High-Dimensional Space*
A slight variation of standard notation allows investigation of a space that accommodates all of the degrees of freedom relevant to a particular problem, together with the ability to focus in on low-dimensional spaces as they become salient. The mathematics serves as a moving spotlight that provides variable views of the sensorimotor space, for insight into the structure of sensorimotor processes.

In a mix of standard set and dimensional notations, we introduce the convention that unspecified degrees of freedom continue to exist and may have any values. To see how this works, first consider two-dimensional sets. For example, a closed disc D in a plane P would be denoted D⊂P. One edge E of a square S can be denoted E⊂S. Alternatively, the square could be denoted as the product of two unit intervals $[0,1]\times[0,1]$ and the edge as $[0,0]\times[0,1] \subset [0,1]\times[0,1]$. The combination of notations can similarly be used for a moving line within a square, a moving rectangle within a square, or for subspaces of spaces of any dimension. But note that the disc can not be written as a product.

Two subsets moving according to different functions of time result in a moving intersection. For example, the moving cube in Fig. 1 is the intersection of two rectangular solids, $[j_x(t,x,y)-\delta, j_x(t,x,y)+\delta]\times[j_y(t,x,y)-\delta, j_y(t,x,y)+\delta]\times[k(t)-\delta, k(t)+\delta] \subset [0,1]\times[0,1]\times[0,1]$, where $j_x, j_y$ are the separate coordinate components, $\delta \leq 0.5$, and $j:(t,x,y)\rightarrow(x,y)$ and $k:(t)\rightarrow(z)$ with $j: \Re^3 \rightarrow [\delta,1-\delta]\times[\delta,1-\delta]$ and $k: \Re \rightarrow [\delta,1-\delta]$. Although $[j_x(t,x,y)-\delta, j_x(t,x,y)+\delta]\times[j_y(t,x,y)-\delta, j_y(t,x,y)+\delta]$ could easily be separated into rectangular components, there are cases in which it is important not to separate a set into rectangular components. For example, in a case in which the motions in the x and y directions form a two-dimensional figure such as a circle, it is more illuminating to draw them as a two-dimensional figure.

The notation accommodates many possibilities that have not been illustrated. For example, the functional relationships can be conditional and discrete, and the directions of the coordinates can be explicitly specified to change as a function of time and other coordinates.

Expressing the time dimension includes explicit dynamics. For example, consider a slice of a cube moving from low to high, with z coordinate specified by a sigmoid curve (Fig. 2). The cube is now explicitly included in a four-dimensional space, including time. The cube itself is a subset, with time suppressed, that is, left unspecified. Because time is unspecified, the included cube as drawn on the left is formally equal to the entire space.

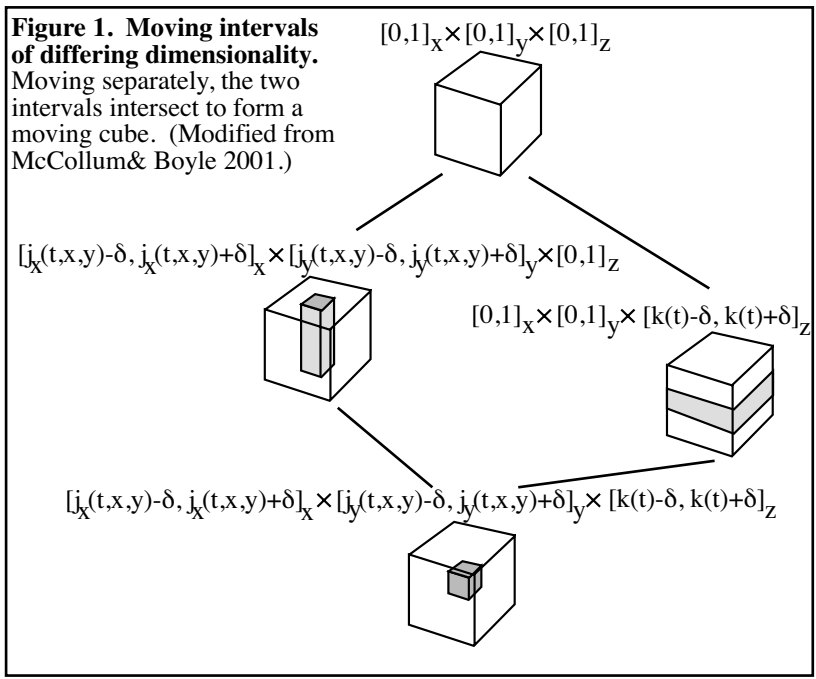

Figure 1. Moving intervals of differing dimensionality. Moving separately, the two intervals intersect to form a moving cube. (Modified from McCollum & Boyle 2001.)

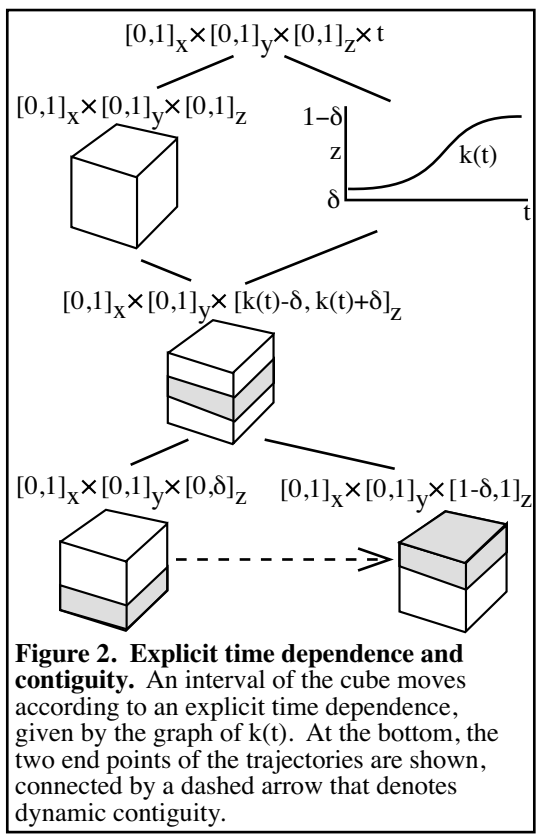

Figure 2. Explicit time dependence and contiguity. An interval of the cube moves according to an explicit time dependence, given by the graph of $k(t)$. At the bottom, the two end points of the trajectories are shown, connected by a dashed arrow that denotes dynamic contiguity.

The graph of $k(t)$, specifying the dynamics of the moving slice, is drawn in z×t space, with the x and y dimensions suppressed. Because of the suppressed degrees of freedom, the moving slice within the cube, just below, is related by bijective correspondence with the set denoted by the graph. The same set can be viewed in different ways, as a 4-, 3-, 2-, or 1-dimensional set.

Further detail is specified at the bottom of Fig. 2 by depicting regions along the path of the moving slice. The slice at the initial position (bottom left) is included within the time sequence of moving slices. The slice at final position is also included (bottom right). The relation of contiguity connects the two positions, as denoted by a dashed arrow. Contiguity depends on the dynamics specified above by the graph of $k(t)$ in z×t space.

The contiguity relation is used to specify regions that occur in the course of dynamics.  Dynamics may be conditionally applicable to the system.  The use of inclusion and contiguity allows the specification of the regions in state space in which dynamics are applicable, along with the results.  The Bloch theorem in dynamics (Bloch, 1995) provides the mathematical justification that suitable subsets can be bracketed for separate dynamical investigation.  The use of algebraic relations -- inclusion and contiguity -- provide a mathematical structure we call "conditional dynamics".

*Structures in Conditional Dynamics*

The use of conditional dynamics allows sensorimotor behaviors to be diagrammed modularly, for example in a clinical setting, and also to be analyzed mathematically, to gain insight into the intrinsic structures of sensorimotor function.  A current issue in neural control is command versus distributed control.  Deterministic or command concepts are congenial under the prevalent mechanistic paradigm and because of the nature of experimental research.  However, insofar as each organism is necessarily autonomous, its decisions must ultimately occur as a state distributed among participating neurons and neural centers.

The simplest autonomous control structure specifying transitions between states along with the conditions for transition is a *dyad* connecting two states (Fig. 3A) (McCollum, 1999b, 2002).  Within each state, the conditions for transition occur, just as preparations for sleep occur during waking.

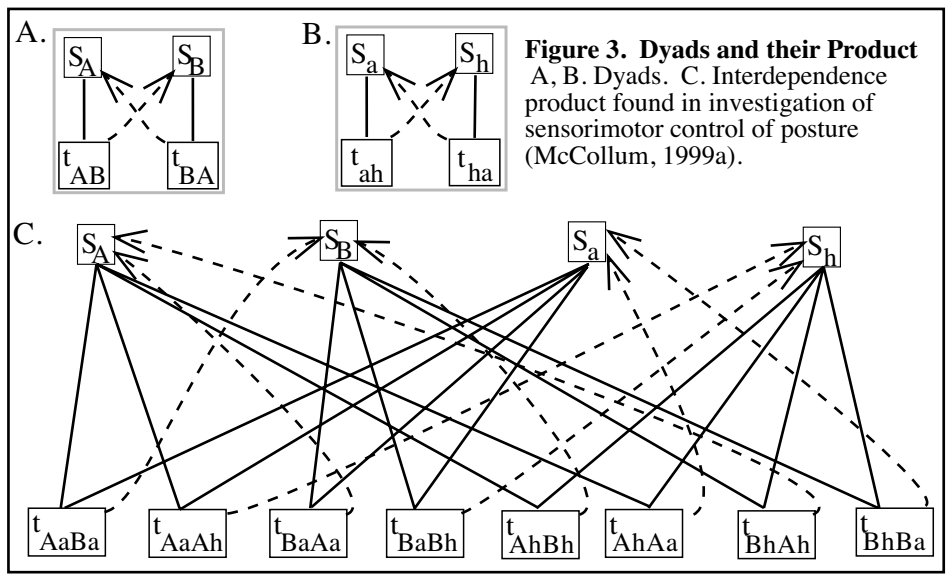

Figure 3.  Dyads and their Product
A, B. Dyads.  C. Interdependence product found in investigation of sensorimotor control of posture (McCollum, 1999a).

In the sensorimotor control of posture, a product between dyads occurs (Fig. 3C) (McCollum, 1999a,b, 2002).  This product can be formalized in various ways.  In current research, we are seeking further empirical examples; the mathematics will be more useful if it is more broadly based within neurobiology.

These structures that have arisen in studies of conditional dynamics in neurobiology suggest a general mathematics of control structures, including both command and autonomous control (McCollum & Boyle, 2001). Current research is aimed at formalizing and generalizing the mathematics of control structures, with mathematical research intertwining with neurobiological research.

*Emergence of Mathematics from Neurobiology*

The generalization of sensorimotor control structures is expected to provide a mathematics related to that of ordered structures. Other mathematical structures are also beginning to be determined in neurobiology, such as physiological groups (Schöner et al 1990; Collins & Stewart, 1992,3; Golubitsky et al, 1998; McCollum & Boyle, 2001/2002, 2002). Although theoretical neurobiology is naturally a subset of neurobiology, as theoretical physics is of physics, mathematicians and physicists have an important role to play in the development of theoretical neurobiology. Physicists have experience in developing mathematics for insight into empirical systems, and mathematicians in the twentieth century have provided a wealth of mathematical structures from which to draw inspiration.

So far, the mathematics we have developed has both drawn on standard concepts in mathematics and extended them to fit the structures found in sensorimotor neurobiology. Within neurobiology, theory synthesizes, taking inspiration from one area of experimental investigation and spreading it to other areas. Just where in mathematics theoretical neurobiology will make its contribution is unpredictable; in any case, the intricate organization of an empirical science is a fertile source of mathematical structure.

**Acknowledgments**
I appreciate the critique and technical assistance provided by Dr. Douglas Hanes and the support from the NIH/NIDCD by grant DC04794.

**References**

Bianchi L, Angelini D, Lacquaniti F. (1998) Individual Characteristics of Human Walking Mechanics. Pflügers Archiv - European Journal of Physiology 436(3): 343-56

Bloch WL (1995) Extending Flows from Isolated Invariant Sets Ergodic Theory and Dynamical Systems 15: 1031-1043

Borghese NA, Bianchi L, Lacquaniti F (1996) Kinematic Determinants of Human Locomotion Journal of Physiology 494(3) 863-79

Collins JJ, Stewart IN (1992). Symmetry-Breaking Bifurcation: A Possible Mechanism for 2:1 Frequency-Locking in Animal Locomotion. Journal of Mathematical Biology 30: 827-838

Collins JJ, Stewart IN (1993). Coupled Nonlinear Oscillators and the Symmetries of Animal

Gaits. Nonlinear Science 3: 349-392

Golubitsky M, Stewart I, Buono P-L, Collins JJ (1998) A Modular Network for Legged Locomotion Physica D 115 56-72

McCollum G (1994a) Dissonance: A Nervous System Analogue to Quantum Incompatibility International Journal of Theoretical Physics 33 41-52

McCollum G (1994b) Navigating a Set of Discrete Regions in Body Position Space Journal of Theoretical Biology 167 263-271

McCollum G (2002) Mathematics Reflecting Sensorimotor Organization. Biological Cybernetics, to appear

McCollum G (1999a) Mutual Causality and the Generation of Biological Control Systems International Journal of Theoretical Physics 38(12): 3253-3267

McCollum G (1999b) Sensory and Motor Interdependence in Postural Adjustments Journal of Vestibular Research 9: 303-325

McCollum G (1994) Navigating a Set of Discrete Regions in Body Position Space Journal of Theoretical Biology 167: 263-271

McCollum G, Boyle R (2001) Conditional Transitions in Gaze Dynamics: Role of Vestibular Neurons in Eye-Only and Eye/Head Gaze Behaviors. Biological Cybernetics 85(6):423-36

McCollum G, Boyle R (2001/2002) Rotations in a Vertebrate Setting: Group Theoretic Analysis of Vestibulocollic Projections. Abstract for the Vestibular Influences on Movement Meeting, Orcas Island, Washington, September 22-26, 2002 Journal of Vestibular Research 11(3-5): 195-196

McCollum G, Boyle R (2002) Rotations in a Vertebrate Setting: Group Theoretic Analysis of Vestibulocollic Projections. submitted for publication

McCollum G, Horak FB, Nashner LM (1985) Parsimony in Neural Calculations for Postural Movements. In J. Bloedel, J. Dichgans, & W. Precht (Eds.) Cerebellar Functions New York, NY: Springer-Verlag

Nashner LM, McCollum G (1985) The Organization of Human Postural Movements: A Formal Basis and Experimental Synthesis The Behavioral and Brain Sciences 8 135-172

Roberts PD, McCollum G (1996) Dynamics of the Sit-to-Stand Movement Biological Cybernetics 74: 147-157

Schöner G, Jiang WY, Kelso JAS. (1990). A Synergetic Theory of Quadrupedal Gaits and Gait Transitions. <u>Journal of Theoretical Biology 142</u>: 359-391